\documentclass[aps,twocolumn,superscriptaddress]{revtex4-1}

\usepackage{amsmath,amssymb,amsfonts,color,graphicx,tabularx}

\usepackage[unicode=true,colorlinks=true]{hyperref}

\hypersetup{linkcolor=blue,citecolor=blue,urlcolor=blue}

\newcommand{\overbar}[1]{\overline{\mkern-2.2mu#1\mkern-0.2mu}}

\begin{document}

\title{Chiral topological states in Bose-Fermi mixtures}

\author{Ying-Hai Wu}
%\email{yinghaiwu88@hust.edu.cn}
\affiliation{School of Physics, Huazhong University of Science and Technology, Wuhan 430074, China}
\affiliation{Max-Planck-Institut f{\"u}r Quantenoptik, Hans-Kopfermann-Stra{\ss}e 1, 85748 Garching, Germany}

\begin{abstract}
Topological states were initially discovered in solid state systems and have generated widespread interest in many areas of physics. The advances in cold atoms create novel settings for studying topological states that would be quite unrealistic in solid state systems. One example is that the constituents of quantum gases can be various types of bosons, fermions, and their mixtures. This paper explores interaction-induced topological states in two-dimensional Bose-Fermi mixture. We propose a class of topological states which have no fractionalized excitations but possess maximally chiral edge states. For previously known topological states, these two features can only be found simultaneously in the integer quantum Hall states of fermions and the $E_{8}$ state of bosons. The existences of some proposed states in certain continuum and lattice models are corroborated by exact diagonalization and density matrix renormalization group calculations. This paper suggests that Bose-Fermi mixture is a very appealing platform for studying topological states.
\end{abstract}

\maketitle

\section{Introduction}
\label{Intro}

The adventure of topological states began in the 1980's with the observation of quantum Hall effect in high quality two-dimensional electron gases (2DEGs)~\cite{Klitzing1980,Tsui1982}. The interest on this topic was further boosted by the discovery of topological insulators in the past decade~\cite{Hasan2010,QiXL2011-1}. These states are so named because they have topology related properties that are insensitive to local perturbations, which is in sharp contrast to the Landau theory of symmetry breaking phases distinguished by local order parameters. It is desirable to classify topological states based on their physical properties and the conditions needed to stabilize them. This is a formidable task that has been partially accomplished in a few cases.

Two fundamental criteria in classification are fractionalization and symmetry protection. If a topological state has elementary excitations carrying fractional quantum numbers of the underlying constituents, it is called fractionalized. If a topological state can only be stable when certain symmetry constraints are satisfied, it is called symmetry-protected. Integer quantum Hall (IQH) states and time-reversal symmetric topological insulators (TIs) are typical examples of non-fractionalized topological states that can be realized using free fermions. The latter ones require protection of time-reversal symmetry and particle number conservation but the former ones do not. Topological states of free fermions under a variety of symmetry conditions have been fully uncovered~\cite{Schnyder2008,Kitaev2009,ChiuCK2016}. One can get a much richer set of phenomena with interactions: topological states that are distinct in free fermions may be connected adiabatically~\cite{Fidkowski2011,WangC2014,YouYZ2014}; topological states that do not exist in free fermions may be enabled~\cite{WangC2017,WuYH2017-2}; topological states may arise in some systems composed of bosons~\cite{Kitaev2011,ChenX2012,LuYM2012}. Fractional quantum Hall (FQH) states are paradigmatic examples of interaction-induced fractionallized topological states in which the elementary excitations carry fractional charges and obey fractional braid statistics. In addition to the fermionic FQH states that have been studied extensively in 2DEGs, bosonic ones are also well-established in theory and actively pursued in experiments~\cite{Regnault2003,ChangCC2005,Gemelke2010}.

One characteristic signature of topological states is edge states. For IQH states and TIs, their bulk topology is characterized by the Chern number and the $Z_2$ invariant defined on closed manifolds. Topological non-triviality would lead to gapless edge states when such states are placed on open manifolds. The bulk-boundary correspondence goes beyond free fermions and also plays an essential role in many fractionalized states~\cite{WenXG1995}. The stability of a topological state can be analyzed by studying the robustness of its edge states. IQH states have chiral edge states where boundary excitations only propagate along one direction and backscattering is forbidden. TIs have helical edge states where boundary excitations can propagate along both directions, so time-reversal symmetry and particle number conservation must be imposed to make sure that no backscattering between the counter-propagating modes would occur. The presence of fractionalized excitations can also prevent helical edge states from being gapped out in certain cases~\cite{Levin2013,Barkeshli2013-2}.

The special properties of topological states make them useful in some technological applications. The precisely quantized Hall conductance of IQH states provides an excellent unit of electrical conductance. The robust edge states may be the foundation of next-generation electronic circuits. The elementary excitations that obey non-Abelian braid statistics in certain systems may serve as qubits in quantum computation to defy errors in a topologically robust way~\cite{Nayak2008}. The exotic properties and potential applications of topological states have inspired the search for them in cold atoms, photons, and superconducting circuits~\cite{Goldman2016,Roushan2017,Ozawa2018,Cooper2018}. There have been tremendous advances in cold atoms since the observation of Bose-Einstein condensation. It is possible to perform experiments on various types of bosons and fermions with Hamiltonians that can be tailored to a great extent. The time-, position-, energy-, and momentum-resolved measurement techniques can help us to extract information in an unprecedented manner.

Topological states of electrons in solid state systems depend crucially on gauge fields, but the analogs of such gauge fields do not appear naturally in the Hamiltonians for cold atoms. This calls for ingenious methods to synthesize effective gauge fields~\cite{Dalibard2011,Goldman2014}. For an atom under rotation, the Coriolis force in the rotating frame has the same effect as a magnetic field. This method is conceptually simple, but a sufficiently strong magnetic field (i.e., high rotation frequency) has not been achieved. Another route that has lead to great success in the past few years is laser-assisted coupling, which can generate many types of gauge fields in continuum and optical lattices. An impressive achievement relevant to this work is the realization of the Harper-Hofstadter model that describes the motion of particles in the presence of both magnetic field and periodic potential~\cite{Harper1955,Hofstadter1976,Aidelsburger2013,Miyake2013,Aidelsburger2015,Mancini2015,Stuhl2015,TaiEM2017}.

This paper considers the novel setting of Bose-Fermi mixture, which has been successfully prepared in cold atoms~\cite{Truscott2001,Modugno2002} but very difficult to find in solid state systems. The prospect of finding topological states in such systems have not been discussed before and we take an important first step in this unexplored territory. It should be emphasized that the bosons and fermions are microscopically independent particles in our systems, which is fundamentally different from the scenario where fermions pair up to form bosons~\cite{Tikofsky1996,YangK2008,LiouSF2017,Repellin2017}. For solid state systems, spin models emerge when the charge degree of electrons is frozen due to strong correlation. It is known that some quantum spin liquids are equivalent to bosonic topological states because spins can be recasted as interacting bosons~\cite{ZhouY2017}. If a system include both itinerant electrons and localized spins, it might be interpreted as a Bose-Fermi mixture, but the interactions between electrons and spins are very different from what will be considered below.

We propose a class of topological states which have no fractionalized excitations but possess maximally chiral edge states (i.e., all branches of the edge states propagate along the same direction). It is perhaps surprising that there are not many non-fractionalized topological states with maximally chiral edge states. Besides the well-known IQH states of free fermions, the only other example is the $E_8$ state of interacting bosons~\cite{Kitaev2011,LuYM2012}~\footnote{It is almost certain that there are other bosonic states with these two properties, but they have not been discussed explicitly in the literature to the best of our knowledge.}. The rarity of non-fractionalized chiral topological states is an intriguing question that deserves further investigation. The classification of such states is somewhat challenging because there are different opinions about whether they should be called short-range or long-range entangled~\cite{Kitaev2011,ChenX2012,LuYM2012}. The rest of this paper is organized as follows. In Sec.~\ref{Model}, continuum and lattice models of Bose-Fermi mixture in synthetic gauge fields are defined. In Sec.~\ref{State}, topological states for these models are proposed and analyzed using wave functions and field theory. In Sec.~\ref{Result}, numerical results are presented to demonstrate that some proposed states do exist in these models. In Sec.~\ref{Conclude}, we conclude the paper with an outlook. The appendices provide some technical details and additional numerical results.

\section{Models}
\label{Model}

The models describe two-dimensional (2D) Bose-Fermi mixture in synthetic gauge fields which act independently on bosons and fermions. The indices $\sigma,\tau=b,f$ will be used as superscripts or subscripts on many symbols to indicate bosons and fermions, respectively. The numbers of particles are denoted as $N^{\sigma}$. 

\subsection{Continuum Model}
\label{ModelA}

\begin{figure}
\includegraphics[width=0.48\textwidth]{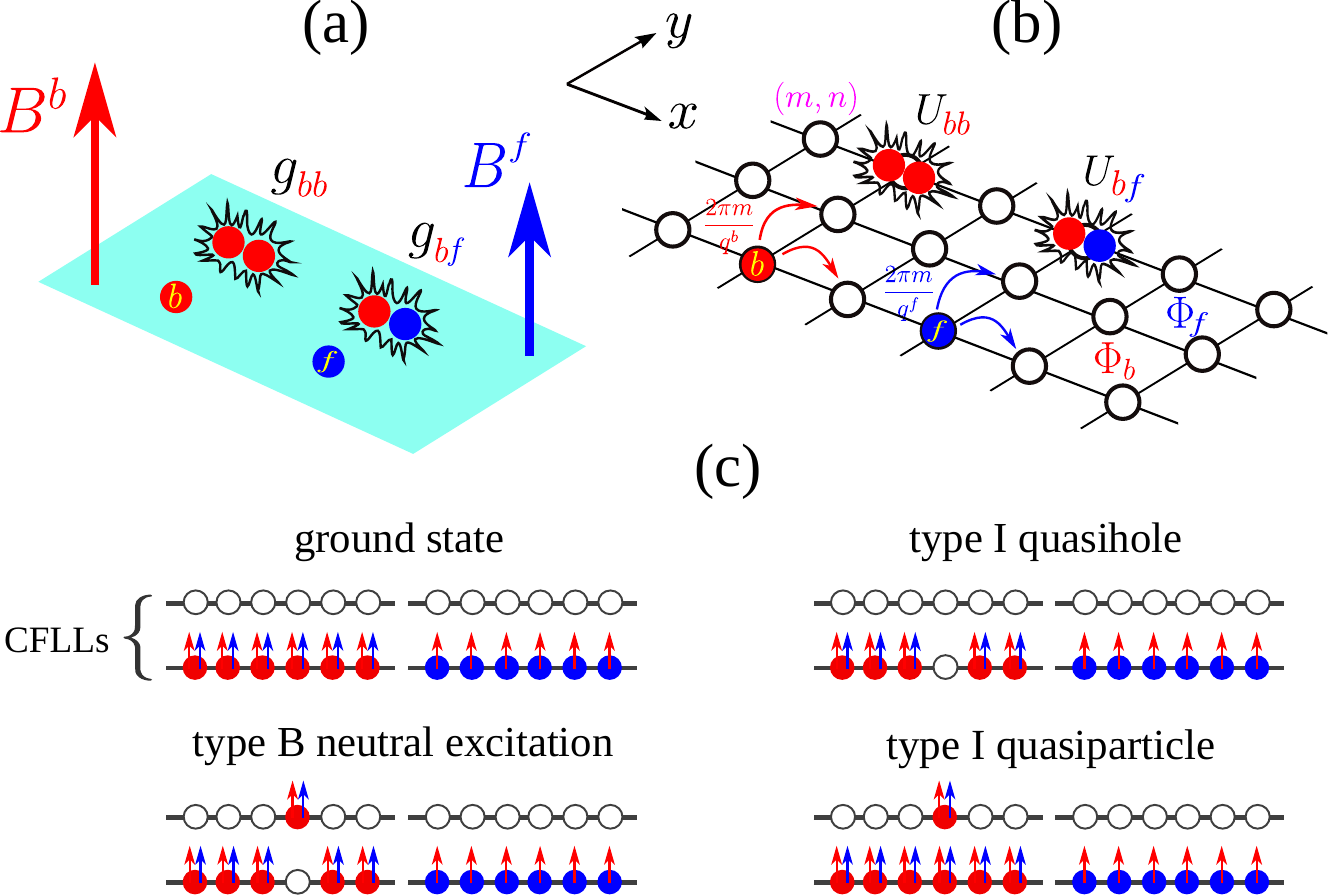}
\caption{(a) Schematics of the continuum model. The bosons (red) and the fermions (blue) experience two independent magnetic fields. The bosons interact with each other via a contact potential of strength $g_{bb}$. The bosons interact with the fermions via a contact potential of strength $g_{bf}$. (b) Schematics of the lattice model. The hopping strengths along the $x$ direction are real and those along the $y$ direction are complex with phases depending on the $x$ coordinates. This gives rise to a phase $\Phi_{b}$ ($\Phi_{f}$) when a boson (fermion) encircles a plaquette. The bosons interact with each other via a onsite interaction of strength $U_{bb}$. The bosons interact with the fermions via a onsite interaction of strength $U_{bf}$. (c) Schematics of the wave function Eq.~\ref{211WaveFunc} and some excitations, where CFLLs stands for composite fermion Landau levels and the arrows represent fluxes attached to particles.}
\label{Figure1}
\end{figure}

As shown in Fig.~\ref{Figure1} (a), the particles experience two independent magnetic fields $B^{\sigma}$ generated by synthetic gauge potentials ${\mathbf A}^{\sigma}$. The single-particle Hamiltonian for the $\sigma$ particles is
\begin{eqnarray}
H^{\sigma}_{0} = \frac{1}{2M^{\sigma}} \left( {\mathbf p} - {\mathbf A}^{\sigma} \right)^2
\label{SingleHamiltonContinuum}
\end{eqnarray}
where $M^{\sigma}$ is the mass of the particles. The solutions to these Hamiltonians are Landau levels (LLs). The specific forms of the single-particle eigenstates depend on the choices of gauge and boundary condition. For our purposes, it is useful to employ disk, torus, and sphere~\cite{Yoshioka1983,Haldane1983}. For an infinite disk with symmetric gauge ${\mathbf A}^{\sigma}=(-B^{\sigma}y/2,B^{\sigma}x/2,0)$, the lowest LL (LLL) wave functions are $\psi^{\sigma}_{m}(x,y) \sim z^{m}$ ($z=x+iy$ is the complex coordinate in two spatial dimensions). The analyticity of these wave functions make them convenient to use in theoretical studies. The bulk properties of a system can be seen more clearly on torus and sphere as they are free of edges. In both cases, the numbers of magnetic fluxes will be quantized and denoted as $F^{\sigma}$. 

The particles interact with each other via the contact interactions
\begin{eqnarray}
\sum_{j<k} \left[ g_{bb} \delta^{bb}({\mathbf r}_{j}-{\mathbf r}_{k}) + g_{bf} \delta^{bf}({\mathbf r}_{j}-{\mathbf r}_{k}) \right]
\label{ManyHamiltonContinuum}
\end{eqnarray}
and we assume that they only occupy their respective LLLs. The second quantized many-body Hamiltonian is
\begin{eqnarray}
V = \frac{1}{2} \sum_{\sigma,\tau} \sum_{\{m_{i}\}} V^{\sigma\tau\tau\sigma}_{m_{1,2,3,4}} C^{\dagger}_{{\sigma}m_{1}} C^{\dagger}_{{\tau}m_{2}} C_{{\tau}m_{4}} C_{{\sigma}m_{3}}
\label{ManyHamiltonContinue}
\end{eqnarray}
where $C^{\dagger}_{\sigma,m}$ ($C^{\dagger}_{\sigma,m}$) is the creation (annihilation) operator for the single-particle state with quantum number $m$ in the LLL of the $\sigma$ particles. The interaction strengths are chosen to be $g_{bb}=g_{bf}=4\pi\ell^{2}_{b}$ such that the zeroth order Haldane pseudopotential is 1 \cite{Haldane1983} and will be used as the unit of energy. The explicit forms of $V^{\sigma\tau\tau\sigma}_{m_{1,2,3,4}}$ on torus, sphere, and disk are given in Appendix~\ref{AppendixA}.

\subsection{Lattice Model}
\label{ModelB}

As shown in Fig.~\ref{Figure1} (b), the particles reside on a square lattice with gauge fields encoded in some complex hoppings of the Harper-Hofstadter type. The numbers of lattice sites along the two directions are denoted as $N_{x}$ and $N_{y}$. The single-particle Hamiltonian is
\begin{eqnarray}
H_{0} &=& - \sum_{\sigma,m,n}  t^{\sigma}_{x} \left( C^\dagger_{\sigma,m+1,n} C_{\sigma,m,n} + {\rm H.c.} \right) \nonumber \\
  && - \sum_{\sigma,m,n} t^{\sigma}_{y} \left( e^{i\phi^{\sigma}_{m}} C^\dagger_{\sigma,m,n+1} C_{\sigma,m,n} + {\rm H.c.} \right)
\label{SingleHamiltonLattice}
\end{eqnarray}
where $C^{\dagger}_{\sigma,m,n}$ ($C^{\dagger}_{\sigma,m,n}$) is the creation (annihilation) operator for the $\sigma$ particles on the lattice site labeled by $m$ and $n$. The hopping coefficients along the $y$ direction contain phases $\phi^{\sigma}_{m}=2{\pi}m/q^{\sigma}$ that depend on the particle species and the $x$ coordinate. The presence of these phases change the translational symmetry of the lattice. It is useful to introduce magnetic unit cells for the two species separately and super magnetic unit cells for the whole system: one bosonic (fermionic) magnetic unit cell includes $q^{b}$ ($q^{f}$) plaquettes along the $x$ direction, one super magnetic unit cell includes ${\rm LCM}(q^{b},q^{f})$ (the least common multiple of $q^{b}$ and $q^{f}$) plaquettes along the $x$ direction, and all of them include one plaquette along the $y$ direction. The model is translationally invariant with respect to the super magnetic unit cell, so periodic boundary condition (PBC) can only be imposed if $N_{x}$ is a multiple of ${\rm LCM}(q^{b},q^{f})$. The particles interact with each other via the onsite interactions
\begin{eqnarray}
V &=& U_{bb} \sum_{m,n} C^\dagger_{b,m,n} C^\dagger_{b,m,n} C_{b,m,n} C_{b,m,n} \nonumber \\
  &+& U_{bf} \sum_{m,n} C^\dagger_{b,m,n} C^\dagger_{f,m,n} C_{f,m,n} C_{b,m,n} 
\label{ManyHamiltonLattice}
\end{eqnarray}
where the two terms correspond to Bose-Bose interaction and Bose-Fermi interaction (there is no on-site Fermi-Fermi interaction).

\section{Topological States}
\label{State}

In this section, we propose a class of topological states and analyze their properties in detail using wave functions in the continuum and effective field theory. The experiences accumulated when studying FQH states provide valuable guides here. The continuum wave functions can be transformed to lattices~\cite{QiXL2011-2,WuYL2012-2,WuYL2013}, but the results are not as easy to analyze as those in the continuum. 

\subsection{Wave Function}
\label{StateA}

The analyticity of the LLL single-particle wave functions on disk geometry makes it ideal for constructing many-body wave functions. It is known that contact interactions and its derivatives often have exact zero-energy eigenstates when they are projected to the LLL~\cite{Haldane1983,Trugman1985}. If Eq.~\ref{ManyHamiltonContinuum} were to have any zero-energy eigenstates, their wave functions must vanish when the distance between any two particles goes to zero (all other particles are kept at fixed positions), otherwise the energy expectation value would be infinite. For all states that satisfy this condition,
\begin{eqnarray}
\prod^{N^{b}}_{j<k} (z^{b}_{j}-z^{b}_{k})^2 \prod^{N^{f}}_{j<k} (z^{f}_{j}-z^{f}_{k}) \prod^{N^{b}}_{j} \prod^{N^{f}}_{k} (z^{b}_{j}-z^{f}_{k})
\label{211WaveFunc}
\end{eqnarray}
has the highest density so it will be taken as the ground state. The first part has a power $2$ instead of $1$ because $\prod^{N^{b}}_{j<k} (z^{b}_{j}-z^{b}_{k})$ does not obey Bose statistics even though the vanishing condition is met. The maximal single-particle angular momentum of bosons is $2N^{b}+N^{f}-2$ and that of fermions is $N^{b}+N^{f}-1$, so the magnetic fields $B^{\sigma}$ should be chosen independently. 

The factor $\prod^{N^{b}}_{j<k} (z^{b}_{j}-z^{b}_{k})^2$ in Eq.~\ref{211WaveFunc} is the bosonic Laughlin $1/2$ state~\cite{Laughlin1983}, which may be replaced by the composite fermion states at filling factor $\mu/(\mu+1)$ ($\mu\in{\mathbb N}$)~\cite{ChangCC2005} (the Laughlin $1/2$ state is reproduced at $\mu=1$) to construct a class of states 
\begin{eqnarray}
\Psi^{b}_{\frac{\mu}{\mu+1}}\left(\{z^{b}\}\right) \prod^{N^{f}}_{j<k} (z^{f}_{j}-z^{f}_{k}) \prod^{N^{b}}_{j} \prod^{N^{f}}_{k} (z^{b}_{j}-z^{f}_{k})
\label{ManyWaveFunc1}
\end{eqnarray}
To express these wave functions compactly, we define $\chi_{\mu}$ as the fermionic IQH state at filling factor $\mu$ [$\chi_{\mu}=\prod^{N^{f}}_{j<k} (z^{f}_{j}-z^{f}_{k})$]. The first part is
\begin{eqnarray}
\Psi^{b}_{\frac{\mu}{\mu+1}}\left(\{z^{b}\}\right) = {\mathcal P} \chi_{\mu}\left(\{z^{b}\}\right) \prod^{N^{b}}_{j<k} (z^{b}_{j}-z^{b}_{k})
\end{eqnarray}
where ${\mathcal P}$ is the LLL projection operator and Eq.~\ref{ManyWaveFunc1} becomes
\begin{eqnarray}
{\mathcal P} \chi_{\mu}\left(\{z^{b}\}\right) \chi_{1}\left(\{z^{f}\}\right) \prod^{N^{b}}_{j<k} (z^{b}_{j}-z^{b}_{k}) \prod^{N^{b}}_{j} \prod^{N^{f}}_{k} (z^{b}_{j}-z^{f}_{k})
\label{ManyWaveFunc2}
\end{eqnarray}
The physical picture for these states is provided by the composite fermion theory~\cite{Jain1989-1}: $\prod^{N^{b}}_{j<k} (z^{b}_{j}-z^{b}_{k})$ dresses each boson with one flux from the other bosons; $\prod^{N^{b}}_{j} \prod^{N^{f}}_{k} (z^{b}_{j}-z^{f}_{k})$ dresses each boson (fermion) with one flux from the fermions (bosons); the flux-attached composite fermions form IQH states $\chi_{\mu}\left(\{z^{b}\}\right)$ and $\chi_{1}\left(\{z^{f}\}\right)$ in effective magnetic fields.

The composite fermion interpretation goes beyond the ground states. Since the composite fermions are taken as non-interacting objects, the excitations of the IQH states $\chi_{\mu}\left(\{z^{b}\}\right)$ and $\chi_{1}\left(\{z^{f}\}\right)$ would give us the excitations of Eq.~\ref{ManyWaveFunc2}. This is the case for both bulk and edge excitations as illustrated schematically in Fig.~\ref{Figure1} at $\mu=1$. If one composite fermion in $\chi_{\mu}\left(\{z^{b}\}\right)$ [$\chi_{1}\left(\{z^{f}\}\right)$] is excited from an occupied LL to an empty LL, one type B (type F) neutral excitation is created. If the magnetic flux for bosons is increased (decreased) by one unit, one type I quasihole (quasiparticle) is created. Type II quasihole and quasiparticle are defined similarly when the magnetic flux for fermions changes. The $\chi_{\mu}\left(\{z^{b}\}\right)$ factor contributes $\mu$ edge modes and the $\chi_{1}\left(\{z^{f}\}\right)$ factor contributes 1 edge modes, so we have a total of $\mu+1$ edge modes.

The charges of quasiholes and quasiparticles deserve special attention. As the gauge potentials for bosons and fermions are independent, we should introduce two kinds of charges for the particles. The bosons (fermions) have unit (zero) charge with respect to ${\mathbf A}^{b}$ and zero (unit) charge with respect to ${\mathbf A}^{f}$. The quasihole charges with respect to ${\mathbf A}^{\sigma}$ are denoted as $Q^{\sigma}_{\rm I}$ and $Q^{\sigma}_{\rm II}$. To find out their values, we study the consequences of removing one boson or fermion from the ground state. If one boson is removed, $\mu+1$ type I quasihole and one type II quasihole are created. If one fermion is removed, $\mu$ type I quasihole and one type II quasihole are created. In the meantime, the total charge of the system with respect to ${\mathbf A}^{\sigma}$ decreases by one unit or stays the same. These facts lead to the equations
\begin{eqnarray}
&& (\mu+1)Q^{b}_{\rm I} + Q^{b}_{\rm II} = -1 \;\;\; {\mu}Q^{b}_{\rm I} + Q^{b}_{\rm II} = 0 \nonumber \\
&& (\mu+1)Q^{f}_{\rm I} + Q^{f}_{\rm II} = 0 \;\;\; {\mu}Q^{f}_{\rm I} + Q^{f}_{\rm II} = -1  
\end{eqnarray}
so we have $Q^{b}_{\rm I}=-1$, $Q^{b}_{\rm II}=\mu$, $Q^{f}_{\rm I}=1$, and $Q^{f}_{\rm II}=-(\mu+1)$. By analyzing the cases with one boson or fermion added to the ground state, one can show that the quasiparticle charges are just opposite to the quasihole charges.

\subsection{Field Theory}
\label{StateB}

The topological field theory for Eq.~\ref{ManyWaveFunc2} is the Chern-Simons theory with Lagrangian density~\cite{WenXG1992-1}
\begin{eqnarray}
{\mathcal L}_{1} = \frac{1}{4\pi\hbar} \epsilon^{\lambda\mu\nu} K_{IJ} a_{I\lambda} \partial_{\mu} a_{J\nu} - j_{I\lambda} a_{I\lambda}
\label{LagrangePartI}
\end{eqnarray}
where $a_{I\lambda}$ ($I=1,2$, $\lambda=0,x,y$) are gauge fields, $j_{I\lambda}$ is the excitation current, and $K$ is an integer matrix. For $\mu=1$ and $2$, the $K$ matrix is
\begin{eqnarray}
K = \left(
\begin{array}{cc}
2 & 1 \\
1 & 1
\end{array}
\right)
\end{eqnarray}
and
\begin{eqnarray}
K = \left(
\begin{array}{ccc}
2 & 1 & 1 \\
1 & 2 & 1 \\
1 & 1 & 1
\end{array}
\right)
\end{eqnarray}
respectively. In general, $K$ is a $(\mu+1)$-dimensional matrix in which $\mu$ diagonal elements are $2$, $1$ diagonal element is $1$, and all off-diagonal elements are $1$. When the system is defined on a torus, the number of degenerate ground states is $|{\det K}|$. For topologically ordered states, we have $|{\det K}|>1$ because fractionalized excitations lead to multiple degenerate ground states. If one subtracts the rightmost column from all other columns, it becomes obvious that the $K$ matrices for all $\mu$ have unit determinant, so these states do not possess fractionalized excitations.

If a system is described by ${\mathcal L}_{1}$ in the bulk, it would have gapless edge modes when placed on an open manifold. The edge physics is captured by the Lagrangian density
\begin{eqnarray}
{\mathcal L}_{\rm edge} = \frac{1}{4\pi\hbar} \left( K_{IJ} \partial_{0} \phi_{I} \partial_{x} \phi_{J} - V_{IJ} \partial_{x} \phi_{I} \partial_{x} \phi_{J} \right)
\end{eqnarray}
where $\phi_{I}$ is a chiral boson field and $V_{IJ}$ depends on the microscopic details at the edge. The propagating directions of the edge modes are determined by the signs of the eigenvalues of the $K$ matrix. The edge states of our system are maximally chiral in the sense that all the eigenvalues of the $(\mu+1)$-dimensional $K$ matrix are positive. This means that the edge states cannot be gapped out by perturbations because there is no back scattering channel.

The charges of quasiparticles and quasiholes can be derived by coupling the excitation currents to two probing gauge fields $\overbar{{\mathbf A}^{b}}$ and $\overbar{{\mathbf A}^{f}}$. This operation adds the Lagrangian density
\begin{eqnarray}
{\mathcal L}_{2} = - \frac{\epsilon^{\lambda\mu\nu}}{2\pi\hbar} \left( t^{b}_I \overbar{A^{b}_{\lambda}} \partial_{\mu} a_{I\nu} + t^{f}_I \overbar{A^{f}_{\lambda}} \partial_{\mu} a_{I\nu} \right)
\label{LagrangePartII}
\end{eqnarray}
to Eq.~\ref{LagrangePartI}, where ${\mathbf t}^{b}=(1,\ldots,1,0)^T$ and ${\mathbf t}^{f}=(0,\ldots,0,1)^T$ are termed charge vectors. An excitation is labeled by an integer vector ${\mathbf l}$ and its charge with respect to ${\mathbf A}^{\sigma}$ is $[\mathbf t^{\sigma}]^{T}K^{-1}{\mathbf l}$. If ${\mathbf l}$ is chosen to be $({\pm}1,0,\ldots,0)^{T}$ for type I quasiparticle/quasihole and $(0,\ldots,0,{\pm}1)^{T}$ for type II quasiparticle/quasihole, the values of $Q^{\sigma}_{\rm I,II}$ computed from wave functions can be reproduced. The braid statistics angle of two excitations labeled by ${\mathbf l}_{1}$ and ${\mathbf l}_{2}$ is $\theta_{11}={\pi}[{\mathbf l}_{1}]^{T}K^{-1}{\mathbf l}_{1}$ if ${\mathbf l}_{1}={\mathbf l}_{2}$ and $\theta_{12}=2{\pi}[{\mathbf l}_{1}]^{T}K^{-1}{\mathbf l}_{2}$ otherwise. The inverse of a matrix can be computed from its determinant and adjugate. The facts that a $K$ matrix has unit determinant and integer elements ensure that all elements of $K^{-1}$ are also integers, so there is no fractional braid statistics between any pairs of excitations, as we expect for systems with a non-degenerate ground state on torus.

\section{Numerical Results}
\label{Result}

In this section, we use exact diagonalization and density matrix renormalization group (DMRG)~\cite{White1992,Schollwock2011,Hubig2015} to study the many-body Hamiltonians Eqs.~\ref{ManyHamiltonContinue} and~\ref{ManyHamiltonLattice} numerically. Exact diagonalization is relatively straightforward and will be used in Sections~\ref{ResultA} and~\ref{ResultB}. DMRG is more involved and will only be used in Section~\ref{ResultB}.

\subsection{Continuum Model}
\label{ResultA}

\begin{figure}
\includegraphics[width=0.48\textwidth]{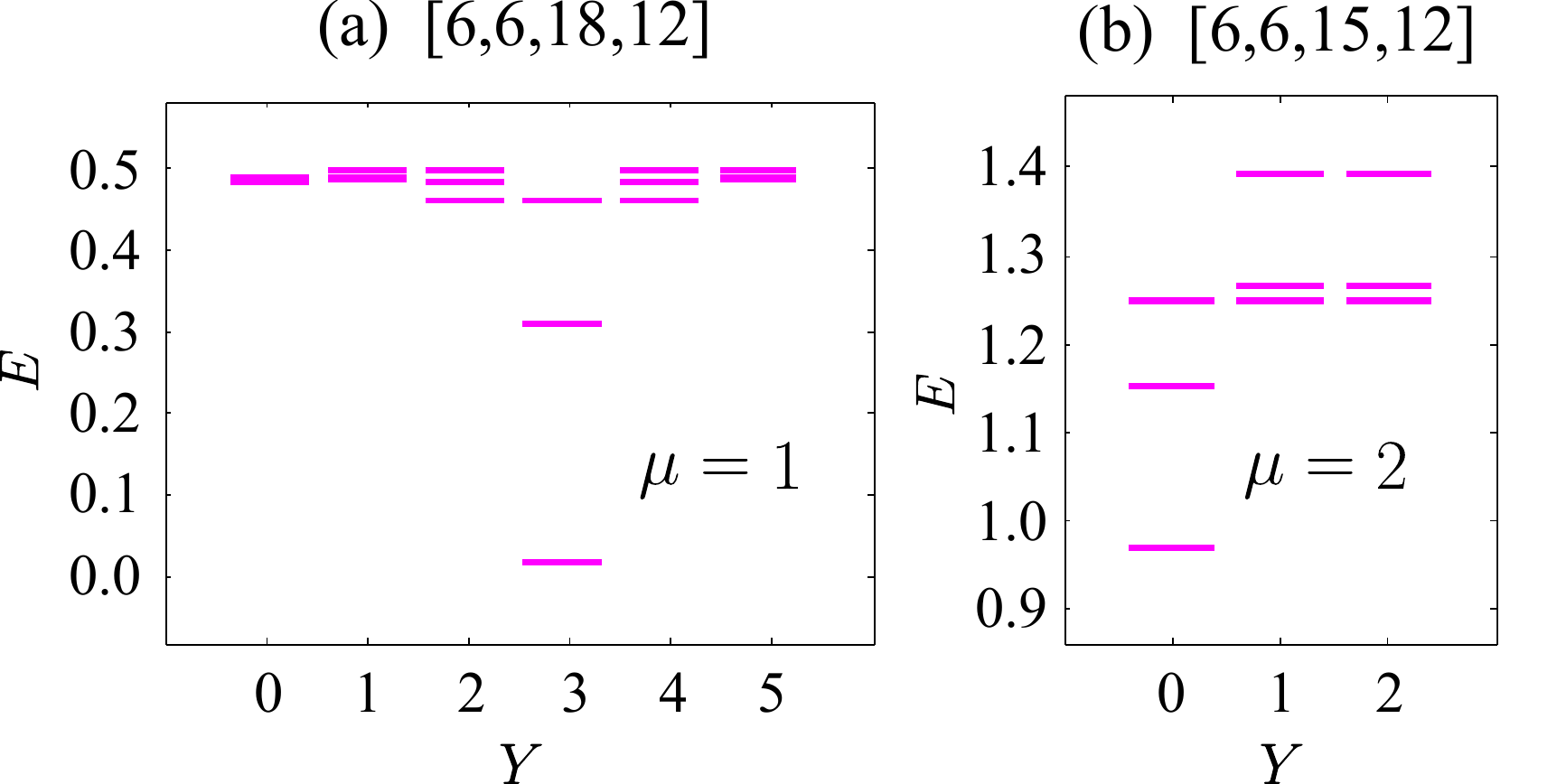}
\caption{Energy spectra of the continuum model on torus. The system parameters are given $[N^{b},N^{f},F^{b},F^{f}]$ in each panel. The energy levels are labeled by the total momentum $Y$. There is a unique ground state in both cases.}
\label{Figure2}
\end{figure}

\begin{figure}
\includegraphics[width=0.48\textwidth]{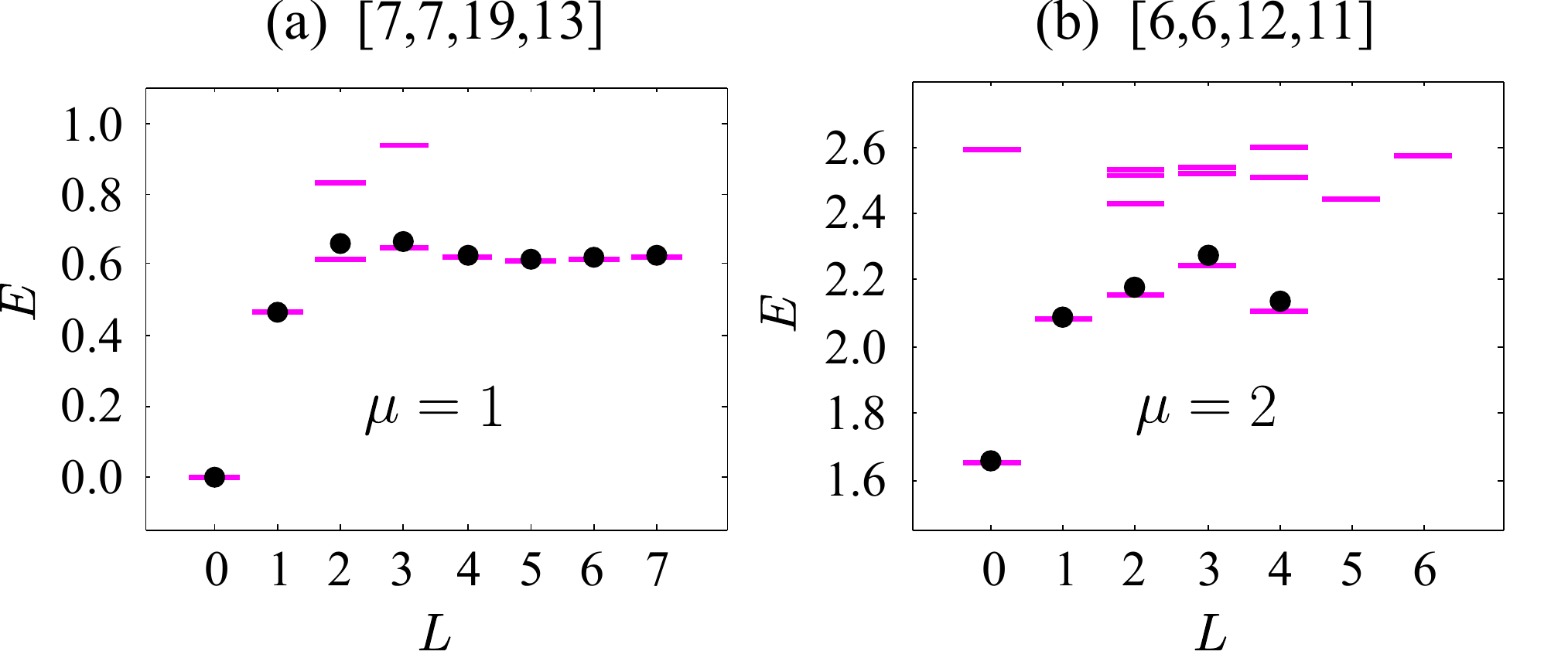}
\caption{Energy spectra of the continuum model on sphere. The system parameters are given $[N^{b},N^{f},F^{b},F^{f}]$ in each panel. The energy levels are the total angular momentum $L$ and its $z$-component $L_{z}$ (chosen to be 0 in both panels). The dots represent counterparts of the wave functions Eq.~\ref{ManyWaveFunc2} on sphere. The overlaps with the exact eigenstates are $1$, $0.9991$, $0.9433$, $0.9882$, $0.9939$, $0.9945$, $0.9939$, $0.9934$ [from $L=0$ to $7$ in (a)] and $0.9982$, $0.9955$, $0.9850$, $0.9796$, $0.9863$ [from $L=0$ to $4$ in (b)].}
\label{Figure3}
\end{figure}

The first signature to be confirmed is the ground state degeneracy on torus. The ground states occur at magnetic fluxes $F^{b}=(\mu+1)N^{b}/\mu+N^{f}$ and $F^{f}=N^{b}+N^{f}$. The Hamiltonian conserves a special momentum $Y$ (see Appendix~\ref{AppendixA} for details). For both $\mu=1$ and $2$, there is a unique ground state on torus as shown in Fig.~\ref{Figure2}, in consistency with the absence of fractionalized excitations. The validity of the wave functions for neutral excitations have been tested on sphere. The ground states occur at magnetic fluxes $F^{b}$=$(\mu+1)N^{b}/\mu+N^{f}-(\mu+1)$ and $F^{f}=N^{b}+N^{f}-1$, where the $O(1)$ quantities $(\mu+1)$ and $1$ are called shifts~\cite{Haldane1983,WenXG1992-2}. The total angular momentum $L$ and its $z$ component $L_{z}$ are good quantum numbers. One may expect that excited composite fermions in either $\chi_{\mu}(\{z^{b}\})$ or $\chi_{1}(\{z^{f}\})$ of Eq.~\ref{ManyWaveFunc2} would result in neutral excitations, but it turns out that the low-energy sector is only made of excited composite fermions in the $\chi_{\mu}(\{z^{b}\})$ factor. For both $\mu=1$ and $2$, the wave functions are very accurate approximations of the exact eigenstates as shown in Fig.~\ref{Figure3}. The ground state energy at $\mu=1$ is indeed zero on torus and sphere. We have also obtained similar results for other available values of $N^{b},N^{f},F^{b},F^{f}$.

\begin{figure}
\includegraphics[width=0.48\textwidth]{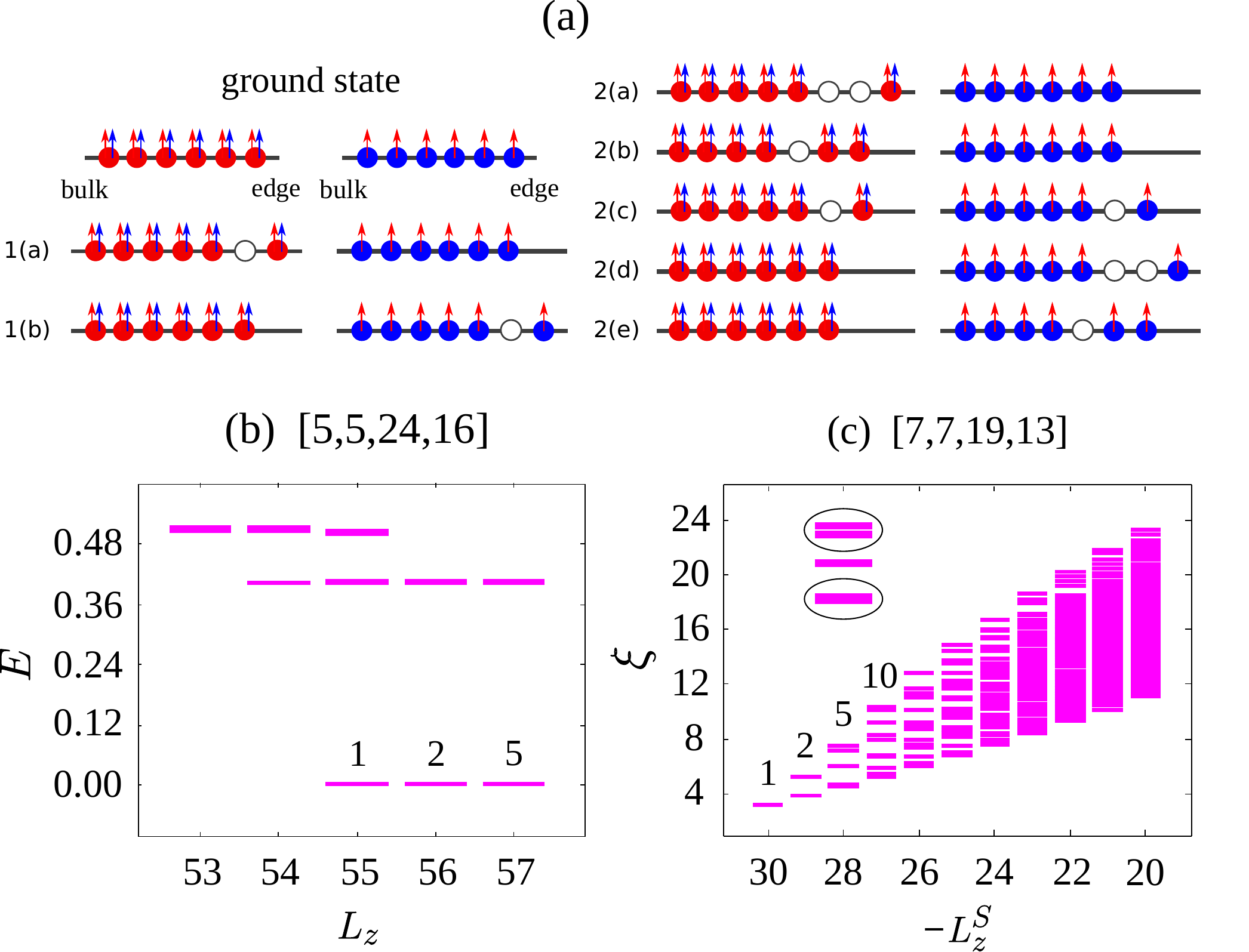}
\caption{Edge physics of the continuum model at $\mu=1$. The system parameters are given as $[N^{b},N^{f},F^{b},F^{f}]$ in each panel. (a) Schematics of the ground state and edge states at ${\Delta L}_z=1,2$. (b) Energy spectrum on disk with the levels labeled by the z-component angular momentum $L_z$. The counting of zero energy states is given in the panel. (c) Entanglement spectrum on sphere with the levels labeled by the $z$-component angular momentum of the southern hemisphere $L^{S}_{z}$. The numbers of particles in the southern hemisphere are $N^{S}_{b}= N^{S}_{f}=4$. The counting of levels is given in the panel. The inset is a zoom-in view where the quasi-degenerate levels at ${\Delta L}_{z}=2$ are circled.}
\label{Figure4}
\end{figure}

The edge states at $\mu=1$ were predicted to consist of two chiral modes. It can be further assumed that they have linear dispersions at low energy. For the disk geometry, the angular momentum and Hamiltonian of the edge excitations are
\begin{eqnarray}
L_{\rm edge} &=& \sum_{m=1} m {\mathcal N}^{\rm I}_{m} + \sum_{m=1} m {\mathcal N}^{\rm II}_{m} \\
H_{\rm edge} &=& v_{\rm I} \sum_{m=1} m {\mathcal N}^{\rm I}_{m} + v_{\rm II} \sum_{m=1} m {\mathcal N}^{\rm II}_{m}
\label{EdgeTheory}
\end{eqnarray}
where ${\mathcal N}^{\rm I,II}_{m}$ are the occupation numbers of the edge modes at angular momentum $m$ (defined relative to the ground state) and $v_{\rm I,II}$ characterize the linear dispersions. For the ground state, we have ${\mathcal N}^{\rm I,II}_{m}=0$ since there is no edge excitation. An important feature of edge physics is how the number of edge states changes with the relative angular momentum $L_{\rm edge}$. By assigning non-zero values to ${\mathcal N}^{\rm I,II}_{m}$, we find that the edge states exhibit a counting $1,2,5,10,\ldots$ as $L_{\rm edge}$ increases from zero (see Table~\ref{Table1} and Fig.~\ref{Figure4} for some examples). This is confirmed in Fig.~\ref{Figure4} by both the energy spectrum on disk and the entanglement spectrum on sphere~\cite{LiH2008}. The edge physics at $\mu=2$ is more complicated and will be discussed in Appendix~\ref{AppendixB}.

\begin{table}
\begin{tabular}{cccc}
\hline
\hline
label & occupation numbers & $L_{\rm edge}$ & $H_{\rm edge}$ \\
\hline
1(a) & ${\mathcal N}^{\rm I}_{m=1}=1$ & 1 & $v_{\rm I}$ \\
1(b) & ${\mathcal N}^{\rm II}_{m=1}=1$ & 1 & $v_{\rm II}$ \\
2(a) & ${\mathcal N}^{\rm I}_{m=2}=1$ & 2 & $2v_{\rm I}$ \\
2(b) & ${\mathcal N}^{\rm I}_{m=1}=2$ & 2 & $2v_{\rm I}$ \\
2(c) & ${\mathcal N}^{\rm I}_{m=1}=1$, f${\mathcal N}^{\rm II}_{m=1}=1$ & 2 & $v_{\rm I}+v_{\rm II}$ \\
2(d) & ${\mathcal N}^{\rm II}_{m=2}=1$ & 2 & $2v_{\rm II}$ \\
2(e) & ${\mathcal N}^{\rm II}_{m=1}=2$ & 2 & $2v_{\rm II}$ \\
\hline
\hline
\end{tabular}
\caption{Edge states at relative angular momentum $1$ and $2$. The states are labeled for reference in the main text. The unspecified occupation numbers are all zero. }
\label{Table1}
\end{table}

There are two issues regarding the energy spectrum on disk. The number of single-particle states is finite on torus and sphere, but is in principle infinite on disk. To get a finite dimensional many-body Hilbert space, we need to introduce cutoffs (also denoted as $F^{\sigma}$) such that the particles can only occupy the states with $m{\leq}F^{\sigma}$. The conservation of the total angular momentum $L_{z}$ gives natural cutoffs, but it is helpful to use smaller values to further reduce the Hilbert space dimension. The energy spectrum in Fig.~\ref{Figure4} (a) is computed using $F^{b}=24$ and $F^{f}=16$, which are sufficiently large in the sense that the first few eigenvalues have almost converged. For instance, if we increase the cutoffs to $F^{b}=25$ and $F^{f}=17$, the variation of the second lowest eigenvalue in the $L_z=55$ sector is less than $10^{-7}$. The existence of multiple degenerate states is a considerable challenge for sparse matrix diagonalization. At angular momentum $3$ relative to the ground state, we can get $10$ zero-energy states for the $N^{b}=N^{f}=4$ system but not for the $N^{b}=N^{f}=5$ system. This issue can be resolved by adding a small perturbation to the Hamiltonian to slightly break the tenfold degeneracy, which results in $10$ quasi-degenerate states well-separated from the other ones.

The entanglement spectrum is an alternative method for probing edge states. For this calculation, the sphere is divided into two hemispheres separated by a virtual edge along its equator~\cite{Dubail2012-1,Sterdyniak2012,Rodriguez2012}. The reduced density matrix $\rho$ of the southern hemisphere is written as $\exp(-H)$. The low-energy part of the entanglement Hamiltonian $H$ captures the edge excitations at the equator. When the eigenvalues $\xi$ of $H$ are plotted versus the good quantum numbers of the southern hemisphere (the numbers of particles and the $z$ component angular momentum), the edge state counting $1,2,5,10$ is revealed. It is instructive to inspect some entanglement levels in greater detail. The relative angular momenta ${\Delta L}_{z}$ of entanglement levels are defined with respect to $L^{S}_{z}=-30$. The edge states have been labeled in Table~\ref{Table1} and their energy values are computed using Eq.~\ref{EdgeTheory}. The differences between the entanglement eigenvalues at ${\Delta L}_{z}=1$ and the one at ${\Delta L}_{z}=0$ give us $v_{\rm I}{\approx}0.68$ and $v_{\rm II}{\approx}2.07$~\footnote{This statement is for the entanglement edge but not the physical edge. The latter is more complicated because $v_{\rm I}$ and $v_{\rm II}$ would depend on the confinement potential and other details at the edge.}. The five levels at ${\Delta L}_{z}=2$ can be divided into three groups: 2(a) and 2(b) are degenerate, 2(d) and 2(e) are degenerate, and 2(c) is separated from others. This is supported by the entanglement spectrum, but the levels are only quasi-degenerate since the dispersions of the entanglement edge modes are not perfectly linear as in Eq.~\ref{EdgeTheory}.

\subsection{Lattice Model}
\label{ResultB}

We first impose PBCs along both directions of the lattice to check the ground state degeneracy. This can be done if the number of sites along the $x$ direction is a common multiple of $q^{b}$ and $q^{f}$. The numbers of energy bands are $q^{\sigma}$ and the numbers of states in each energy band are $N_{x}N_{y}/q^{\sigma}$. In analogy to the continuum LL on torus, we have the relations $N_{x}N_{y}/q^{b}=(\mu+1)N^{b}+N^{f}$ and $N_{x}N_{y}/q^{f}={\mu}N^{b}+N^{f}$. The system parameters need to satisfy two more constraints to stabilize the topological states of our interest. One is that the lowest energy band should be sufficiently flat (so the single-particle contributions are not important) and the other is that $N_{x}/N_{y}$ should not be too large or too small (so the system is not in the thin torus limit~\cite{Bergholtz2005,Seidel2005,Bernevig2012-2}). These conditions can be met using $N^{b}=N^{f}, q^{b}=6, q^{f}=9$ at $\mu=1$, but no suitable parameters are found at $\mu=2$. 

\begin{figure}
\includegraphics[width=0.48\textwidth]{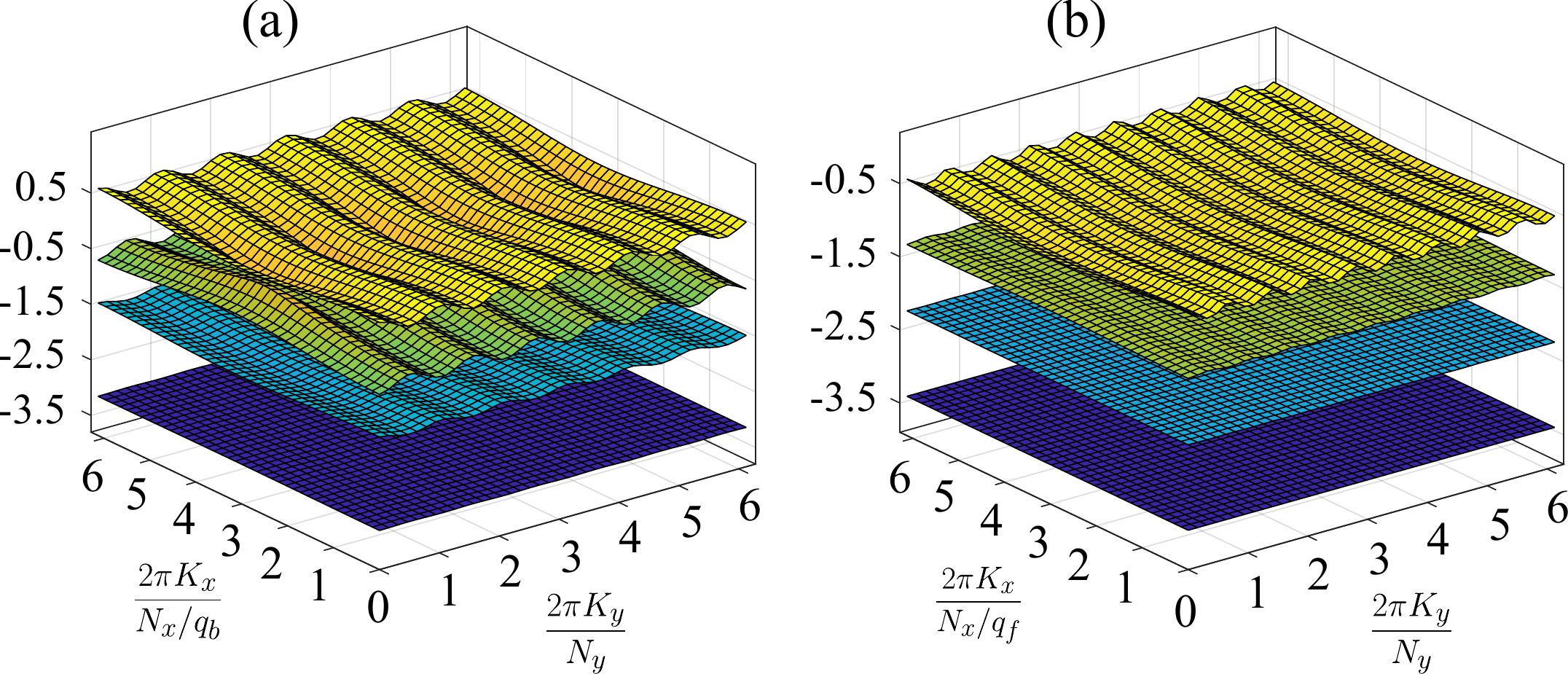}
\caption{Energy bands of the hopping terms in Eq.~\ref{SingleHamiltonLattice}. (a) The bosonic part with $q^{b}=6,t^{b}_{x}=1.0,t^{b}_{y}=1.0$ on the lattice with $N_{x}=240,N_{y}=40$. (b) The fermionic part with $q^{f}=9,t^{f}_{x}=1,t^{f}_{y}=1.0$ on the lattice with $N_{x}=360,N_{y}=40$. There are multiple bands in both cases but only four of them are shown.}
\label{Figure5}
\end{figure}

\begin{figure}
\includegraphics[width=0.48\textwidth]{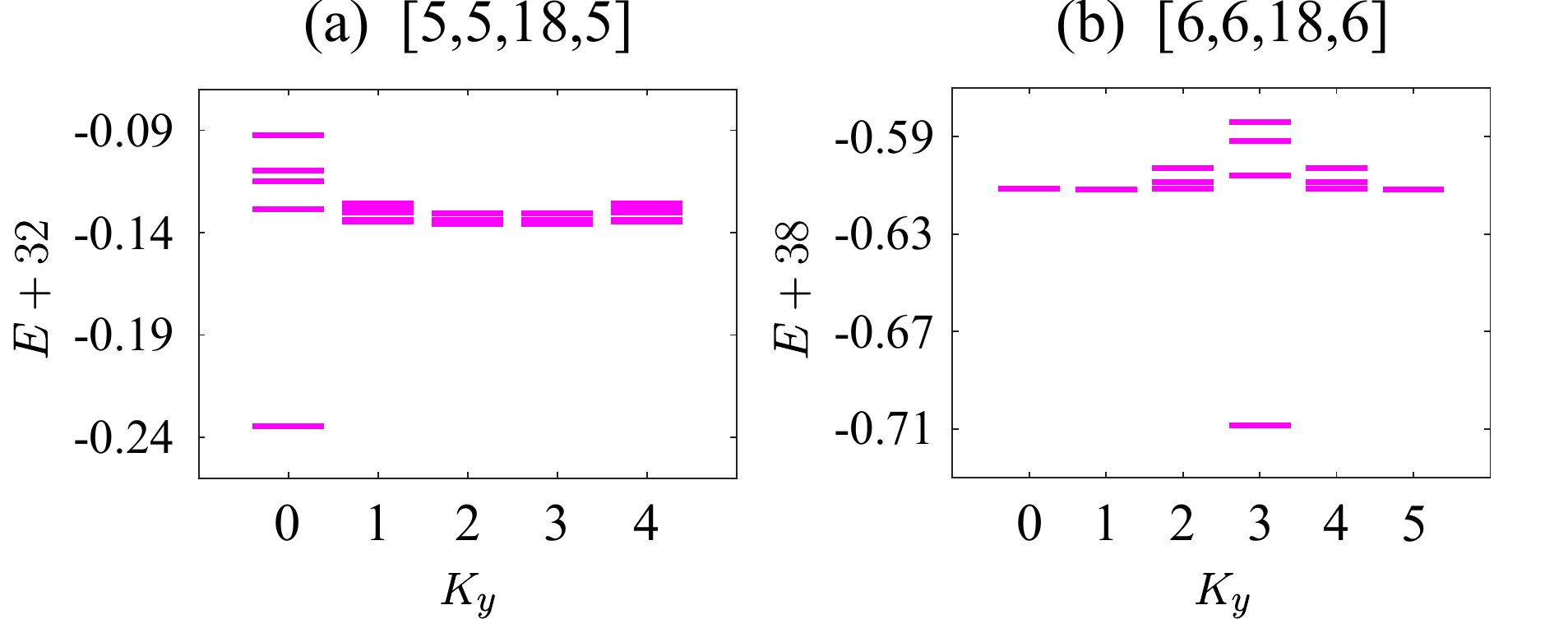}
\caption{Energy spectra of the lattice model at $\mu=1$ on torus. The system parameters are given as $[N^{b},N^{f},N_{x},N_{y}]$ in each panel. The Hamiltonian parameters are $q^{b}=6$, $q^{f}=9$, $t^{b}_{x}=t^{f}_{x}=1.0$, $t^{b}_{y}=t^{f}_{y}=1.0$, $U_{bb}=1$, and $U_{bf}=2$. The energy levels are labeled by the quantum number $K_{y}$ alone since there is only one super magnetic unit cell along the $x$ direction.}
\label{Figure6}
\end{figure}

The full problem contains multiple energy bands so exact diagonalization cannot be performed on any systems with reasonable sizes. To this end, we project the Hamiltonian to the lowest bands of the lattice model in the same spirit as the LLL approximation adopted in the continuum model. The term ``lowest bands" should be clarified here. The bosonic and fermionic hopping terms in Eq.~\ref{SingleHamiltonLattice} give us two independent sets of energy bands (see Fig.~\ref{Figure5}), and the interactions in Eq.~\ref{ManyHamiltonLattice} are projected to the lowest bands of each species. When the particles are treated together in the full system, it is only translationally invariant with repsect to the super magnetic unit cell that contains multiple bosonic (fermionic) magnetic unit cells. This means that the bosonic and fermionic lowest bands are folded to produce some bands for the full system. The energy levels are labeled by the lattice momenta $2{\pi}{\rm LCM}(q_{b},q_{f})K_{x}/N_{x}$ and $2{\pi}K_{y}/N_{y}$ defined with respect to the super magnetic unit cell. The two energy spectra in Fig.~\ref{Figure6} both have a unique ground state.

\begin{figure}
\includegraphics[width=0.48\textwidth]{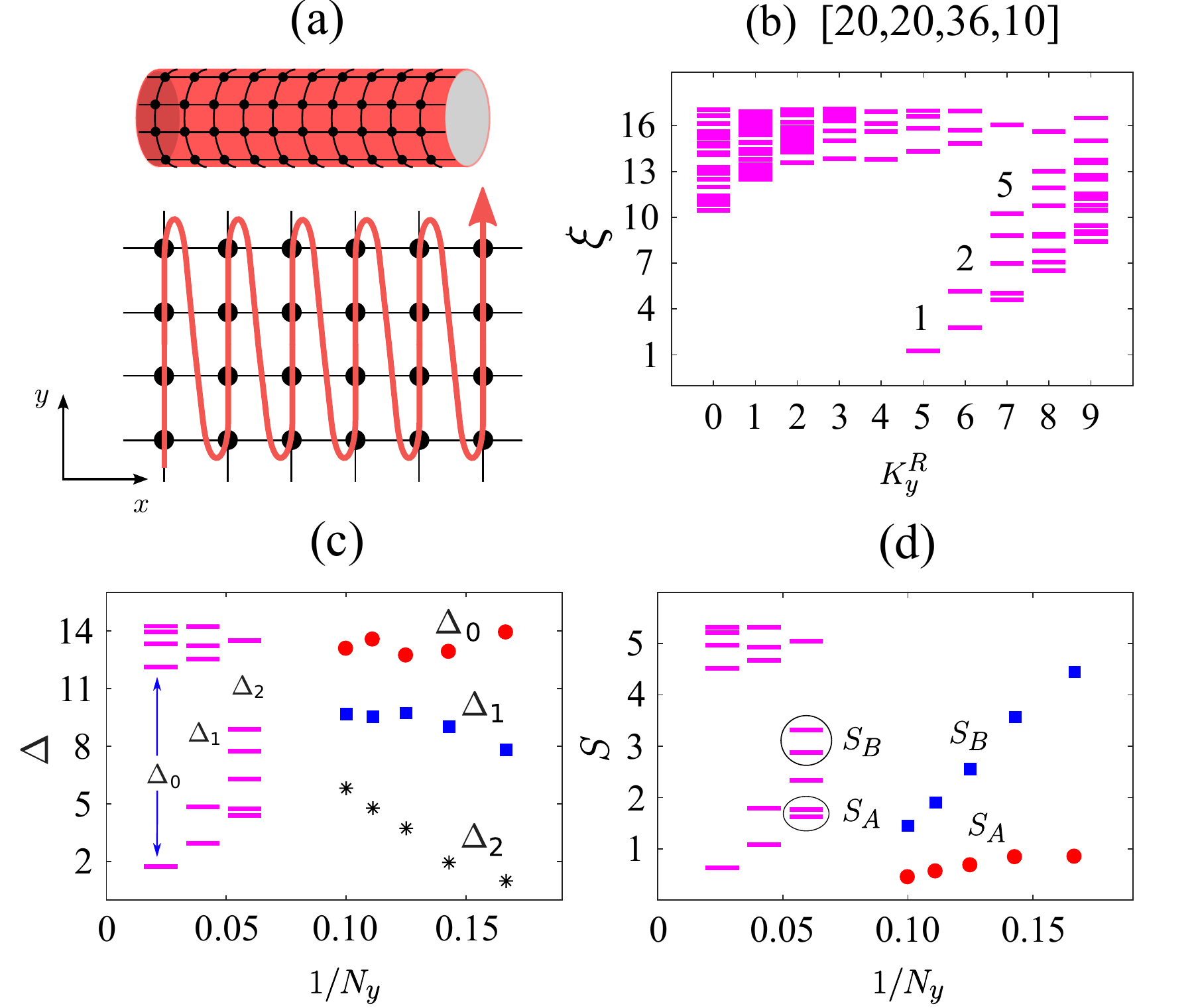}
\caption{DMRG result of the lattice model at $\mu=1$ on cylinder. (a) Schematics of a cylinder and the mapping of a 2D lattice into a 1D chain. (b) Entanglement spectrum on the cylinder with the levels labeled by the total momentum $K^{R}_{y}$ of the right half. The system parameters are given as $[N^{b},N^{f},N_{x},N_{y}]$ in this panel. The Hamiltonian parameters are $q^{b}=6$, $q^{f}=9$, $t^{b}_{x}=t^{f}_{x}=1.0$, $t^{b}_{y}=t^{f}_{y}=1.0$, $U_{bb}=\infty$, and $U_{bf}=5$. The numbers of particles in the right half are $N^{b}_{R}=N^{f}_{R}=10$. The counting of levels is given in the panel. (c) Finite size scaling of the entanglement gaps in panel (b). (d) Finite size scaling of the level splittings in panel (b). The insets of panels (c) and (d) reproduce part of the entanglement spectrum to define the relevant quantities (see main text for more details).}
\label{Figure7}
\end{figure}

The lattice model can be studied using DMRG without lowest band projection. The DMRG algorithm is a variational method within the class of matrix product states. It is designed for one-dimensional (1D) systems and generally has lower computational cost when the boundary is open. To apply DMRG in 2D models, a 2D lattice should be mapped to a 1D chain as in Fig.~\ref{Figure7} (a). This process poses substantial numerical challenge because many short-range hoppings in 2D are converted to long-range hoppings in 1D. The 2D lattice is chosen to be open (periodic) along the $x$ ($y$) direction, so there is no long-range hopping between the two edges of the cylinder. In contrast to exact diagonalization, the hard-core interaction $U_{bb}=\infty$ is adopted in DMRG calculations. This is a very good approximation when $U_{bb}>10$ and a favorable condition for its relatively low computational cost.

Edge states on lattice can be probed using entanglement spectrum. The lattice is divided into left and right halves separated by a virtual edge in the middle. The reduced density matrix of the right half is used to define the entanglement spectrum similar to what was done on sphere. The $1,2,5$ counting appears in Fig.~\ref{Figure7} (b) when the levels are plotted versus the good quantum numbers of the right half (the numbers of particles and the momentum along $y$). The entanglement spectrum on lattice is not as good as that on sphere: there are non-universal levels that do not represent edge states and the degeneracy at $K^{R}_{y}=7$ is not very good. To demonstrate that the edge physics is not disproved by these imperfections, we perform finite size scaling analysis of entanglement gaps in Fig.~\ref{Figure7} (c) and level splittings in Fig.~\ref{Figure7} (d). The entanglement gap $\Delta_{i}$ is the separation between the highest level in the universal part and the lowest level in the non-universal part (with $i=0,1,2$ correspond to the $K^{R}_{y}=5,6,7$ sectors). The level splittings $S_{A,B}$ are the separations between the two pairs of levels in the $K^{R}_{y}=7$ sector that would be degenerate for perfectly linear edge modes. It is diffcult to be very conclusive, but the results suggest that the entanglement gaps remain finite and the level splittings decay to zero as $N_{y}\rightarrow\infty$.

\section{Conclusions}
\label{Conclude}

In summary, we have proposed a class of topological states for 2D Bose-Fermi mixture in synthetic gauge fields. The analysis based on wave functions and field theory reveal that these states have no fractionalized excitations but maximally chiral edge states. For all previously known topological states, these two features only appear simultaneously in the IQH states of free fermions and the $E_{8}$ state of interacting bosons. The existences of some proposed states in simple continuum and lattice models are corroborated by numerical calculations. This paper demonstrates that mixing bosons and fermions can potentially lead to a plethora of topological states. The three parts in Eq.~\ref{ManyWaveFunc1} may be replaced to construct other wave functions: the first part could be non-Abelian bosonic FQH states or composite fermion liquid~\cite{Moore1991,Halperin1993}, the second part could be any fermionic FQH states, and the third part could have another power. One may also use conformal field theory and parton theory to construct Bose-Fermi topological states~\cite{Moore1991,Jain1990}. It would require a considerable amount of effort to properly classify such states. The bosons and fermions were assumed to be fundamentally distinct in our previous discussions. If this constraint is relaxed such that the bosons can be paired fermions, the Bose-Fermi mixed states might be reinterpreted as topological states of interacting fermions. The implication of this finding on the classification of fermionic topological states is still unclear. All these stimulating questions are left for future works.

\section*{Acknowledgement} 

Exact diagonalization calculations are performed using the DiagHam package for which we are grateful to all the authors. This work was supported by the NSFC under Grant No. 11804107, startup grant of HUST, and the DFG within the Cluster of Excellence NIM.

\begin{widetext}

\begin{appendix}

\setcounter{figure}{0}
\renewcommand\thefigure{A\arabic{figure}}

\section{Hamiltonian Matrix Elements}
\label{AppendixA}

The magnetic length for the $\sigma$ component is $\ell_{\sigma}=\sqrt{\hbar/B^{\sigma}}$. The coefficients $V^{\sigma\tau\tau\sigma}_{m_{1}m_{2}m_{4}m_{3}}$ are
\begin{eqnarray}
\int \; d^{2}{\mathbf r}_{1} d^{2}{\mathbf r}_{2} \; \left[ \psi^{F^{\sigma}}_{m_{1}}({\mathbf r}_{1}) \right]^{*} \left[ \psi^{F^{\tau}}_{m_{2}}({\mathbf r}_{2}) \right]^{*} 4\pi\ell^2_{b} \delta({\mathbf r}_{1}-{\mathbf r}_{2}) \; \psi^{F^{\tau}}_{m_{4}}({\mathbf r}_{2}) \psi^{F^{\sigma}}_{m_{3}}({\mathbf r}_{1})
\end{eqnarray}

\subsection{Torus}

A torus can be constructed from a rectangle by imposing PBCs in both directions. If the torus is spanned by the vectors $\mathbf{L}_{1}=L_{1}{\widehat e}_{x}$ and $\mathbf{L}_{2}=L_{2}{\widehat e}_{y}$, the reciprocal lattice vectors are ${\mathbf G}_{1}=2{\pi}{\widehat e}_{x}/L_{1}$ and ${\mathbf G}_{2}=2{\pi}{\widehat e}_{y}/L_{2}$. In the Landau gauge ${\mathbf A}^{\sigma}=(0,B^{\sigma}x,0)$, the LLL single-particle wave functions on a torus with $F^{\sigma}$ magnetic fluxes are
\begin{eqnarray}
\psi^{F^{\sigma}}_{m} &=& \frac{1}{(\sqrt{{\pi}}L_{2}\ell_{\sigma})^{1/2}} \sum^{\mathbb{Z}}_{k} \exp \Big\{ - \frac{1}{2} \left[ \frac{x}{\ell_{\sigma}} - \frac{2{\pi}\ell_{\sigma}}{L_{2}} \left( m+kF^{\sigma} \right) \right]^2 + i \frac{2{\pi}y}{L_{2}} \left( m+kF^{\sigma} \right) \Big\}
\end{eqnarray}
The magnetic length is related to the lengths of torus via $L_{1}L_{2}=2{\pi}F^{b}\ell^{2}_{b}=2{\pi}F^{f}\ell^{2}_{f}$. The coefficients $V^{\sigma\tau\tau\sigma}_{m_{1}m_{2}m_{4}m_{3}}$ are
\begin{eqnarray}
\frac{1}{F^{b}} \sum_{m_{1}} \sum_{m_{2}} \sum^{{\mathbb Z}}_{q_{1},q_{2}} \exp \left\{ -\frac{{\mathbf q}^2}{4}(\ell^2_{\sigma}+\ell^2_{\tau}) + i2{\pi}q_{1} \left[ \frac{(m_{1}-q_{2}/2)}{F^{\sigma}} - \frac{(m_{2}+q_{2}/2)}{F^{\tau}} \right] \right\} {\widetilde\delta}^{F^{\rm G}}_{m_{1}+m_{2},m_{3}+m_{4}}
\end{eqnarray}
where ${\mathbf q} = q_{1}{\mathbf G}_{1} + q_{2}{\mathbf G}_{2}$, $F^{\rm G}$ is the greatest common divisor of $F^{b}$ and $F^{f}$, and ${\widetilde\delta}^{F^{\rm G}}_{i,j}$ is a generalized Kronecker delta function defined as
\begin{eqnarray}
{\widetilde\delta}^{F^{\rm G}}_{i,j}=1 \;\; {\rm iff} \;\; i \; {\rm mod} \; F^{\rm G} = j \; {\rm mod} \; F^{\rm G}
\end{eqnarray}
The many-body eigenstates are labeled by the total momentum $Y\equiv(\sum_{\sigma={b,f}}\sum^{N^{\sigma}}_{i=1} m^{\sigma}_{i}) \; {\rm mod} \; F^{\rm G}_{\phi}$.

\subsection{Sphere}

A magnetic monopole at the center of a sphere creates a uniform magnetic field through its surface. The LLL single-particle wave functions on a sphere with $F^{\sigma}$ magnetic fluxes are \cite{WuTT1976}
\begin{eqnarray}
\psi^{F^{\sigma}}_{m}(\theta,\phi) = \left[ \frac{F^{\sigma}+1}{4\pi} \binom{F^{\sigma}}{F^{\sigma}-m} \right]^{\frac{1}{2}} u^{F^{\sigma}/2+m} v^{F^{\sigma}/2-m}
\end{eqnarray}
where $\theta$ and $\phi$ are the azimuthal and radial angles, $u=\cos(\theta/2)e^{i\phi/2},v=\sin(\theta/2)e^{-i\phi/2}$ are spinor coordinates, and $m$ is the $z$ component angular momentum. The magnetic length is related to the radius of the sphere via $R=\ell_{b}\sqrt{F^{b}/2}=\ell_{f}\sqrt{F^{f}/2}$. The coefficients $V^{\sigma\tau\tau\sigma}_{m_{1}m_{2}m_{4}m_{3}}$ are
\begin{eqnarray}
\frac{1}{R^2} S_{m_{1}m_{2}} S_{m_{3}m_{4}} \delta_{m_{1}+m_{2},m_{3}+m_{4}}
\end{eqnarray}
where $S_{m_{1}m_{2}}$ and $S_{m_{3}m_{4}}$ are defined by
\begin{eqnarray}
\psi^{F^{\sigma}}_{m_{1}} \psi^{F^{\tau}}_{m_{2}} = (-1)^{F^{\sigma}-F^{\tau}} \psi^{F^{\sigma}+F^{\tau}}_{m_{1}+m_{2}} S_{m_{1}m_{2}} \;\;\; \psi^{F^{\sigma}}_{m_{3}} \psi^{F^{\tau}}_{m_{4}} = (-1)^{F^{\sigma}-F^{\tau}} \psi^{F^{\sigma}+F^{\tau}}_{m_{3}+m_{4}} S_{m_{3}m_{4}} \\
S_{m_{1}m_{2}} = \left[ \frac{(F^{\sigma}+1)(F^{\tau}+1)}{4\pi(F^{\sigma}+F^{\tau}+1)} \right]^{1/2} \left\langle \frac{F^{\sigma}}{2},-m_{1} ; \frac{F^{\tau}}{2},-m_{2} \Bigg| \frac{F^{\sigma}+F^{\tau}}{2},-m_{1}-m_{2} \right\rangle
\end{eqnarray}
The many-body eigenstates are labeled by the total angular momentum $L$ and its $z$ component $L_{z}$.

\subsection{Disk}

The LLL single-particle wave functions on a disk with magnetic length $\ell_{\sigma}$ are
\begin{eqnarray}
\psi^{\sigma}_{m}(x,y) = \frac{z^{m}\exp[-{|z|^2}/(4\ell^{2}_{\sigma})]}{\ell^{m+1}_{\sigma}\sqrt{2{\pi}2^{m}m!}}
\end{eqnarray}
The coefficients $V^{\sigma\tau\tau\sigma}_{m_{1}m_{2}m_{4}m_{3}}$ are
\begin{eqnarray}
\frac{2g^{m_{2}+m_{4}}_{\sigma}g^{m_{1}+m_{3}}_{\tau}}{(g^{2}_{\sigma}+g^{2}_{\tau})^{\sum_{i}(m_{i}/2)+1}} \frac{\Gamma\left[\sum_{i}(m_{i}/2)+1\right]}{\sqrt{\prod^{4}_{i=1}m_{i}!}} \delta_{m_{1}+m_{2},m_{3}+m_{4}}
\end{eqnarray}
where $g_{\sigma,\tau}=\ell_{\sigma,\tau}/\ell_{b}$ and $\Gamma(x)$ is the gamma function.

\setcounter{figure}{0}
\renewcommand\thefigure{B\arabic{figure}}

\section{Edge States at $\mu=2$}
\label{AppendixB}

The edge states at $\mu=2$ is more complicated than the $\mu=1$ case. For the disk geometry, the numbers of states in each LL are infinite. The $\chi_{2}$ state is ambiguous for a finite size system on disk because one has the freedom to decide the numbers of particles in each LL. The true ground state on disk can only be determined after the confinement potential is specified, but we have not studied such cases for simplicity. For each valid finite size representation of $\chi_{2}$, we can construct the wave function Eq.~\ref{ManyWaveFunc2} and compare it with exact diagonalization result. The two examples in Fig.~\ref{FigureB1} demonstrate that the wave functions are accurate approximations of the exact eigenstates.

\begin{figure}
\includegraphics[width=0.7\textwidth]{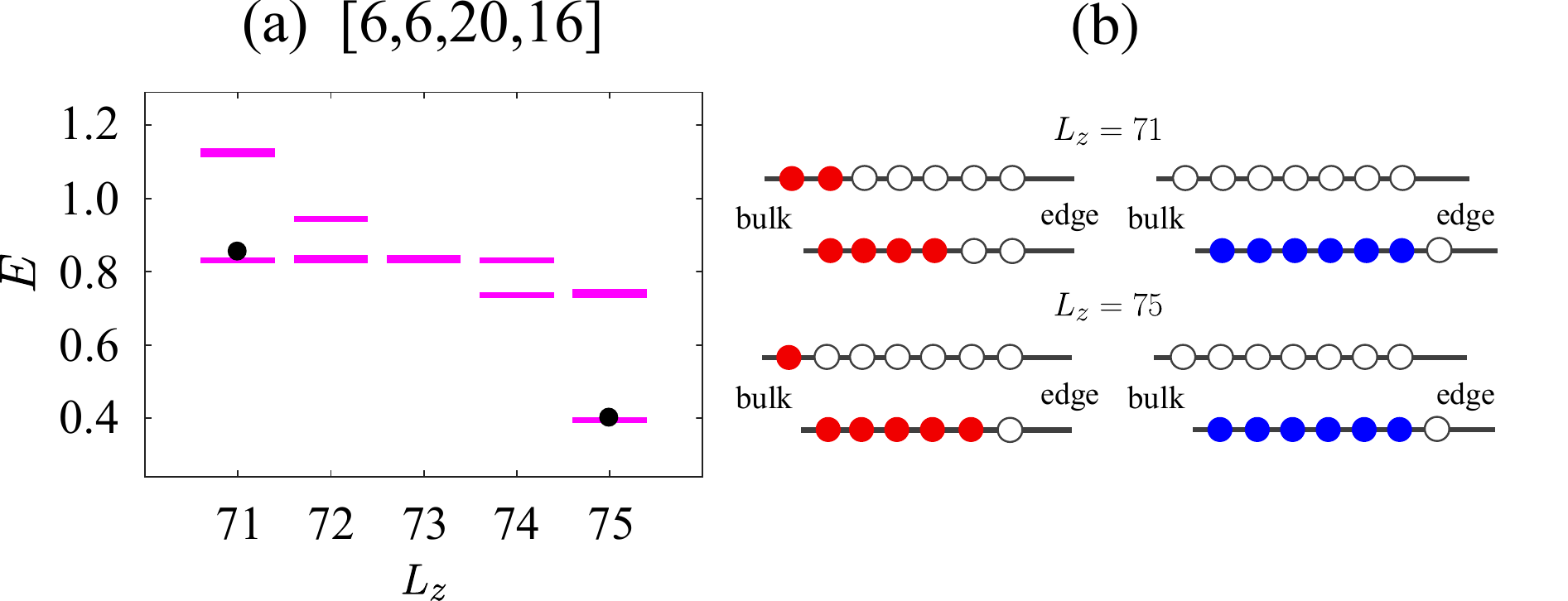}
\caption{(a) Energy spectrum of the continuum model at $\mu=2$ on disk. The system parameters are given as $[N^{b},N^{f},F^{b},F^{f}]$ in each panel. The dots represent the wave functions in Eq.~\ref{ManyWaveFunc2} with their composite fermion configurations given in panel (b).}
\label{FigureB1}
\end{figure}

\end{appendix}

\end{widetext}

\bibliography{../ReferCollect}

\end{document}